# Fluctuating charge density waves in a cuprate superconductor

Darius H. Torchinsky,[1,*] Fahad Mahmood,[1,*] Anthony T. Bollinger,[2] Ivan Božović,[2] and Nuh Gedik[1,†]

[1]Department of Physics, Massachusetts Institute of Technology, Cambridge, Massachusetts, 02139, USA

[2]Brookhaven National Laboratory, Upton, New York 11973, USA

[*]These authors contributed equally

[†]To whom correspondence should be addressed (gedik@mit.edu)

**Cuprate materials hosting high-temperature superconductivity (HTS) also exhibit various forms of charge and/or spin ordering[1-6] whose significance is not fully understood[7]. To date, static charge-density waves (CDWs)[8] have been detected by diffraction probes only at special doping[9-11] or in an applied external field[13]. However, dynamic CDWs may also be present more broadly and their detection, characterization and relationship with HTS remain open problems. Here, we present a new method, based on ultrafast spectroscopy, to detect the presence and measure the lifetimes of CDW fluctuations in cuprates. In an underdoped $La_{1.9}Sr_{0.1}CuO_4$ film ($T_c$ = 26 K), we observe collective excitations of CDW that persist up to 100 K. This dynamic CDW fluctuates with a characteristic lifetime of 2 ps at $T$ = 5 K which decreases to 0.5 ps at $T$ = 100 K. In contrast, in an optimally doped $La_{1.84}Sr_{0.16}CuO_4$ film ($T_c$ = 38.5 K), we detect no signatures of fluctuating CDWs at any temperature, favoring the competition scenario. This work forges a path for studying fluctuating order parameters in various superconductors and other materials.**

Among the rich structural, magnetic, and electronic phases of the cuprates, there exist various forms of modulated charge order[1-6]. A particularly well-studied example is found in $La_2$- $_x$(Sr$_x$,Ba$_x$)CuO$_4$ for $x \sim 0.02 - 0.14$, where a fraction of the carriers forms a charge density wave (CDW)[8] coexistent with magnetic density wave ordering in a configuration sometimes referred to as "stripes"[9,12]. In a narrow doping region around $x = 1/8$, the charge ordering in La$_{2-x}$Ba$_x$CuO$_4$ (LBCO) is static and eliminates superconductivity[13]. However, in La$_{2-x}$Sr$_x$CuO$_4$ (LSCO), the impact of the density wave is somewhat mitigated[14], causing only a slight depression in $T_c$ in the vicinity of $x = 1/8$. This is likely due to the dynamical nature of the density wave, which fluctuates with a finite correlation time $\tau_F$ (represented schematically in Fig 1a).

This fluctuating characteristic has been a central point in the debate surrounding the role of CDWs in the cuprates[7], in particular, whether they favor - or even enable - HTS, or compete with it. In the competition scenario the stability of the fluctuating CDW should decrease with higher values of $T_c$ since both the CDW and the superconducting order would compete for charge carriers. In order to study fluctuating order, researchers have relied on quasi-static probes of local order[7] in a configuration where the density wave has been stabilized either via an external magnetic field[15] or by adding specific dopants, e.g., Nd (Refs. 9, 10) or Eu (Ref. 11). These requirements have only recently been surmounted by the use of resonant inelastic x-ray scattering[5] and high energy x-ray diffraction[6] although these techniques have yet to provide dynamical information about the CDW from which the critical behavior of $\tau_F$ may be elucidated.

In contrast, ultrafast metrologies provide a new opportunity to probe dynamics of charge correlations directly without the need for stabilizing fields or impurities[16]. Among the various degrees of freedom that may be probed are the collective modes of the CDW, as depicted in Fig. 1b. Here, absorption of an ultrafast laser pulse generates single particle excitations of the CDW and hot carriers[17]. These may be manipulated to either coherently drive oscillations of the magnitude of the charge order, i.e., the amplitude mode (amplitudon), or induce the collective sliding of the modulated charge, i.e., the phase mode (phason). The amplitudon is generated by uniform illumination of the sample while the phason may be driven, as we describe below, by a sinusoidally modulated excitation density – a transient grating – created by the interference of two temporally and spatially coincident beams.

In the cuprates, the amplitudon and phason of the CDW have so far been probed via Raman scattering in LSCO single crystals[18, 19] where their broad frequency-domain features did not allow a reliable extraction of their lifetimes. In contrast, a time-domain approach, similar to studies in conventional CDW systems[17, 20-23], would allow an accurate characterization of these highly damped modes, and thus provide information on the dynamical nature of the CDW and yield its fluctuation time $\tau_F$.

Here, using ultrafast spectroscopy, we present evidence for the coherent generation and detection of the collective modes of the fluctuating CDW in a $La_{1.9}Sr_{0.1}CuO_4$ thin film. Following uniform photoexcitation by an ultrashort laser pulse, we observe highly damped oscillations in the reflectivity of a time delayed probe pulse (pump-probe spectroscopy). The frequency of these oscillations as well as temperature and excitation density dependence of their amplitude indicates that they arise from the amplitudon of the CDW. When we perturb the system with a spatially varying sinusoidal excitation density (transient grating spectroscopy[24], see Methods), we observe an additional slow response which we ascribe to the phason. Using predictions for the temperature evolution of the phason damping rate[25, 26], we obtain an estimate of the CDW fluctuation lifetime $\tau_F$.

Figure 2a shows pump-probe (PP) data at a succession of temperatures. For $T < T_c$, there is a short $\sim$ 1 ps "spike" which is identical to the response for $T > T_c$ and was not observed to change in dynamics up to 300 K. We interpret this as arising from uncondensed electrons and will refer to it below as the "normal" component. This fast electronic response is followed at short times by highly damped oscillations which are superposed upon the slow response due to quasiparticle recombination; this slow response has been studied extensively[39, 40] and will not be discussed further here. The oscillation is isolated by subtracting the slow response (Fig. 2a inset). As the temperature is increased across $T_c$, the oscillations persist with decreasing strength up to ~100 K, above which they cannot be discerned from the noise.

In order to study the CDW dynamics without the presence of the superconducting state, we show data taken in the PP and transient grating (TG) geometry at 45 K (i.e., $T > T_c$) (Fig. 2b). Here, we observe the presence of the normal response described above and an additional component which provides a TG response that decays more slowly than that measured in the PP geometry. This additional response disappears when the temperature is raised to 100 K (Fig. 2b inset). In contrast with the data of Fig. 2a and 2b, there are no oscillations in an optimally doped $La_{1.84}Sr_{0.16}CuO_4$ ($T_c$ = 38.5 K) sample at any temperature, as seen in Figure 2c for 5 K and 50 K. We further note that this sample produced a TG response which was identical to the PP response above $T_c$, as shown in Fig. 2d.

We now proceed to establish that the oscillatory response of Fig. 2a originates from the amplitudon. Adapting an approach similar to Demsar et al.[17], we fit the data of Fig. 2a to a model which comprises both the electronic and oscillatory components as $\Delta R(t)/R = A e^{-t/\tau_A}\sin(2\pi f_A t) + Cs(t)$, where $A$ is the magnitude of the oscillating component, $\tau_A$ its lifetime and $f_A$ its frequency. $C$ represents the strength of the electronic response $s(t)$ due to quasiparticle recombination below $T_c$ and due to charge relaxation above $T_c$. A representative fit is shown in Fig. 3a, where we observe excellent agreement between the model function and the data, yielding the frequency $f_A = 2.0$ THz (67 cm$^{-1}$) and the lifetime $\tau_A = 300$ fs. Neither $f_A$ nor $\tau_A$ were observed to vary with temperature within our experimental uncertainties, although the amplitude $A$ of the response diminishes with temperature until it is no longer detectable, as shown in Fig 3c. This value for $f_A$ is lower in frequency than any optical phonon mode observed by frequency-domain Raman technique[27], for which the full theoretical assignment of the lattice modes in the parent material has been made[28]. Rather, $f_A$ approximately matches the value assigned by Sugai et al.[19] to the amplitudon in their study of LSCO single crystals by Raman spectroscopy.

In the absence of optical phonon modes that match the observed oscillation, we conclude that this response is due to an amplitudon driven by a mechanism akin to the displacive excitation of coherent phonons (DECP)[17, 29]. In this scenario, depicted in Fig. 3d, single-particle excitations and hot carriers arising from optical excitations of the CDW alter the local potential $V_0$ of the CDW. Their sudden photo-generation results in an impulsive change of the potential energy landscape of the modulated order $V'$ which shifts the equilibrium charge modulation away from the photoexcited state configuration. In response, the system oscillates about the new equilibrium charge configuration with frequency $f_A$, modulating the reflectivity.

Further evidence to support the identification of the oscillation as the amplitudon is provided in the plot of $A/(\Delta R(0)/R)$, i.e., the ratio of the amplitude (A) of the amplitudon to the total amount of signal ($\Delta R(0)/R$), as a function of the incident laser fluence in Fig. 3b. Here, we observe that the oscillation strength decreases relative to the total amount of signal as the excitation fluence is increased, indicating a saturation of the oscillatory response. While for a Raman-active phonon mode the strength of the phonon response is proportional to the electron dynamics that drive the phonon, in the CDW case the finite concentration of the participating carriers (estimated[19] at ~10%) is depleted by the pump pulse at a relatively modest fluence of ~ 3 $\mu J/cm^2$. We therefore conclude that the oscillation of Fig. 2a is not due to an optical phonon, but specifically to the amplitudon of the CDW. Thus, the temperature at which the amplitudon disappears indicates that $T_{CDW} \approx 100$ K, consistent with observations from Raman spectroscopy[19]. Since the processes responsible for the amplitudon damping rate may possibly be faster than $\tau_F$, the 300 fs lifetime measured above thus sets a lower bound on the CDW fluctuation lifetime.

We now consider the decay component of Fig. 2b observed in the TG configuration. The data were analyzed by first fitting the PP response at 100 K. This produced an excellent phenomenological fit which was subtracted from the raw TG data at all temperatures to account for the unchanging normal component observed by the PP transients. The resulting TG – PP data were fitted to a single exponential $Pe^{-t/\tau_P}$ to represent the overdamped decay. A representative fit is presented in Fig. 4a, showing agreement between the raw data and the model predictions. As with the amplitudon, the magnitude of the overdamped response $P$ diminishes with temperature and recedes below a detectable level at 100 K (Fig. 3c), indicating its common origin with the amplitudon signal.

We have considered other possible sources that can give rise to a difference between the TG and the PP signals, such as propagating optical[30] and acoustic phonons[31] at the wavelength $\Lambda$ of the grating, and thermal[31] or carrier diffusion[32]. None of these responses could cause the TG signal of Fig. 2b: photo-generated coherent acoustic phonons in this system appear at too low frequencies[38] while the optical phonon modes frequencies are too high; both the optic and the acoustic phonon modes are underdamped[38], and the selection rules for optical phonons would imply that the response would be present both in the PP and TG geometries. Thermal diffusion is much slower than the ~ 1 ps timescales observed here, and both thermal and carrier diffusion would produce a TG response that is *faster* than in the PP geometry[32].

By ruling out alternative sources of the TG signal and considering the concomitant disappearance of the amplitudes $A$ and $P$, we conclude that the phason is responsible for the TG signal in Fig. 2b, and is generated via a mechanism, shown in Fig. 4b, that is similar to that which drives the amplitudon. Since the CDW is known to be incommensurate in LBCO and LNSCO even at the $x = 1/8$ doping of static charge order[33, 34], the phason is gapless. Impurities pin the phason and prohibit it from sliding[35] unless a depinning field is applied. However, due to the presence of the same single-particle excitations and hot carriers that drive the amplitudon, there is a change in the local electronic potential $V(x)$ experienced by the CDW as a function of position $x$. This change $\delta V(x)$ shares the spatial profile of the driving laser field, i.e., it is Gaussian in the PP case with the FWHM of the laser, while in the TG geometry it is both Gaussian and spatially periodic with grating period $\Lambda$. In the PP configuration, the Gaussian of FWHM 60 $\mu$m produces a maximal gradient approximately two orders of magnitude smaller than that generated by the typical grating spacing in the TG geometry with grating period $\Lambda = 6$ $\mu$m. The result is an in-plane, spatially periodic electric field $E = -\nabla \delta V(x)$ that serves as the depinning field and drives the phason.

In Fig. 4c, we show the lifetime $\tau_P$ deduced from the fits as a function of temperature. It decreases monotonically from $\tau_P \sim 2$ ps at 5 K to $\tau_P < 500$ fs at 100 K. There are two mechanisms responsible for the lifetime of the phason: intrinsic damping and the lifetime of the CDW fluctuations. Intrinsic damping occurs due to the emission of lower-energy phasons or phonons[25] and yields a damping rate $\Gamma_P = 1/\tau_P$ which has been shown to scale as $T^2$ at low temperatures and as $T^5$ closer to $T_{CDW}$[25, 26, 36]. Plotting the value of $\Gamma_P$ obtained from the fits as a function of temperature in Fig. 4d on a log-log scale, we observe a crossover from sublinear behavior ($\Gamma_P \propto T^{1/4}$) at low temperatures ($T < 40$ K) to linear behavior ($\Gamma_P \propto T$) at higher temperatures ($T > 40$ K), in contrast with theoretical predictions. Moreover, the lifetime of the detected phason does not depend on its wave-vector (see Supplementary Information). These findings rule out intrinsic damping as the only source of the lifetime. We therefore posit that the short lifetime of the phason is due to the disappearance of the CDW within its lifetime $\tau_F$. Since the charge modulation itself is fluctuating, the elementary excitations of the fluctuating order cannot persist longer than the order itself. Measurement of the phason's temporal evolution thus provides direct access to the temporal evolution of the fluctuating CDW.

We have used ultrafast pump-probe and transient grating methods to observe the presence of fluctuating charge order in thin films of $La_{1.90}Sr_{0.10}CuO_4$ via the amplitudon and phason of the CDW. The amplitudon is observable through an oscillation in the PP reflectivity transients $\Delta R(t)/R$, while the phason is manifested through an additional relaxation component in the TG channel. Our data indicate that $T_{CDW} \sim 100$ K and that the fluctuating CDW lifetime varies from $\tau_F = 2$ ps at 5 K to $\tau_F = 500$ fs at 100 K. These experiments provide the first direct dynamical measurement of modulated charge in cuprates and establish ultrafast spectroscopies as a valuable probe of fluctuating CDW order. Absence of these modes in the optimally doped sample (with a higher $T_c$) strongly suggests that fluctuating CDW competes with HTS.

## Methods

Samples of $La_{1.9}Sr_{0.1}CuO_4$ ($T_c$ = 26 K) and $La_{1.84}Sr_{0.16}CuO_4$ ($T_c$ = 38.5 K) were prepared by atomic-layer-by-layer molecular beam epitaxy (ALL-MBE)[37]. Details of sample synthesis and characterization are given in the supplementary information. Experiments were performed with a Ti:sapphire oscillator lasing at the center wavelength of 795 nm ($\hbar\omega$ = 1.55 eV) producing pulses 60 fs in duration. The repetition rate of the laser was reduced to 1.6 MHz with an extracavity pulsepicker to avoid cumulative heating effects on the sample. Two separate experimental geometries were used. In the pump-probe (PP) geometry, the sample was excited by a spatially Gaussian "pump" pulse of 60 $\mu$m FWHM. The sample response was then recorded via measurement of the normalized change in the reflectivity, $\Delta R(t)/R$, of a separate "probe" pulse as a function of delay time $t$ between the pump and the probe. In the transient grating (TG) geometry, the intersection of two equal-intensity, temporally coincident copies of the excitation pulse at an angle $\theta$ produces a sinusoidal modulation of the excitation profile with spacing $\Lambda = \lambda/(2\sin(\theta/2))$, where $\lambda$ is the laser wavelength. Signal was recorded using a beam that both diffracted off the induced grating and reflected specularly from the excited region. The contribution due to the sinusoidal excitation was separated from the transients to yield the TG response[24]. Further details are given in the supplementary information.

## Author Contributions

Experiments were performed by D. H. T. and F. M. The films were synthesized and characterized by A.T.B and I.B. D. H. T. and F. M. performed the data analysis and wrote the initial draft of the manuscript. All authors participated in the understanding of the data and contributed to the final version of the manuscript. N. G. conceived and supervised the project.

The authors declare no competing financial interests.

## Acknowledgements

The authors would like to thank Senthil Todadri, Patrick Lee, and Steve Kivelson for useful discussions. D.H.T, F.M. and N.G. were supported by NSF Career Award DMR-0845296. I.B. and A.T.B were supported by U.S. Department of Energy, Basic Energy Sciences, Materials Sciences and Engineering Division.

**Figure 1**

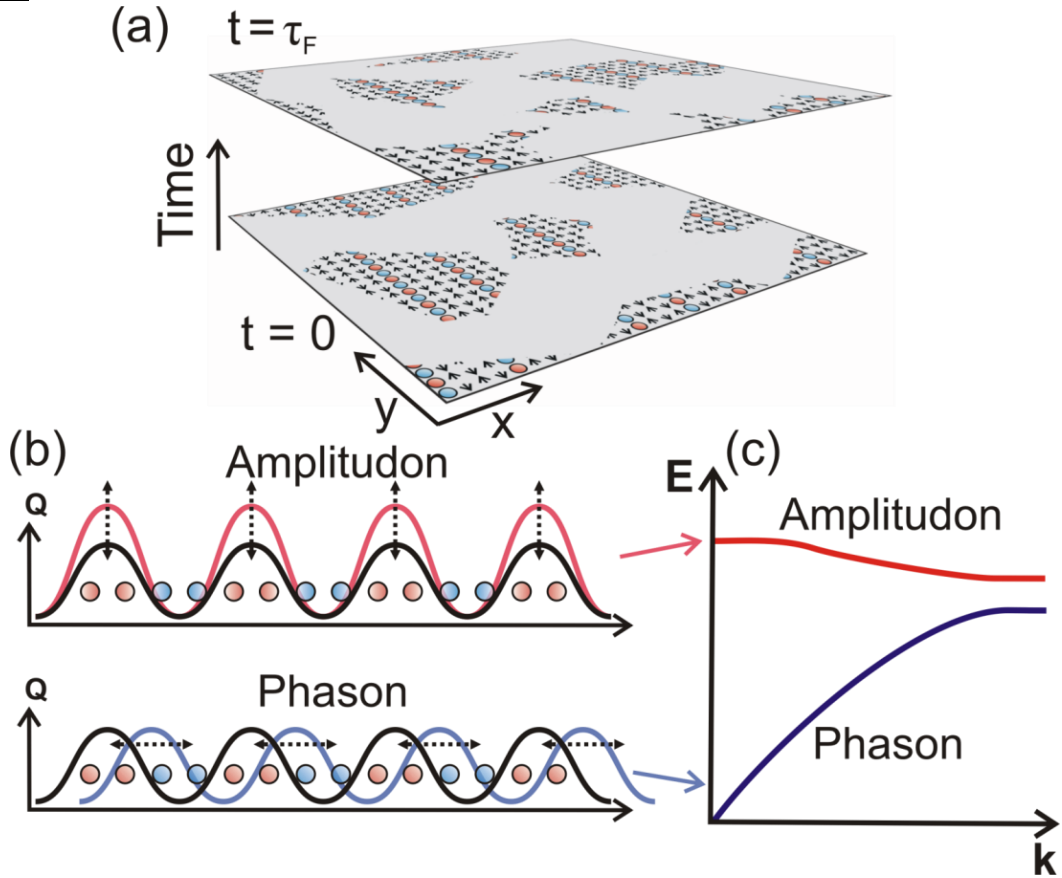

**Figure 1.** Fluctuating density waves and their normal modes. **(a)** This schematic illustrates the dynamics of the fluctuating CDW. The spatial organization of the CDW order changes within the fluctuation lifetime $\tau_F$. **(b)** The CDW gives rise to two collective excitations: the amplitude mode (amplitudon), which represents an overall oscillation of the CDW amplitude, and the phase mode (phason), which is due to a sliding of the CDW along the modulation direction. **(c)** The amplitudon exhibits optical dispersion while the acoustic phason dispersion is gapless due to the incommensurability of the density wave with respect to the lattice.

**Figure 2**

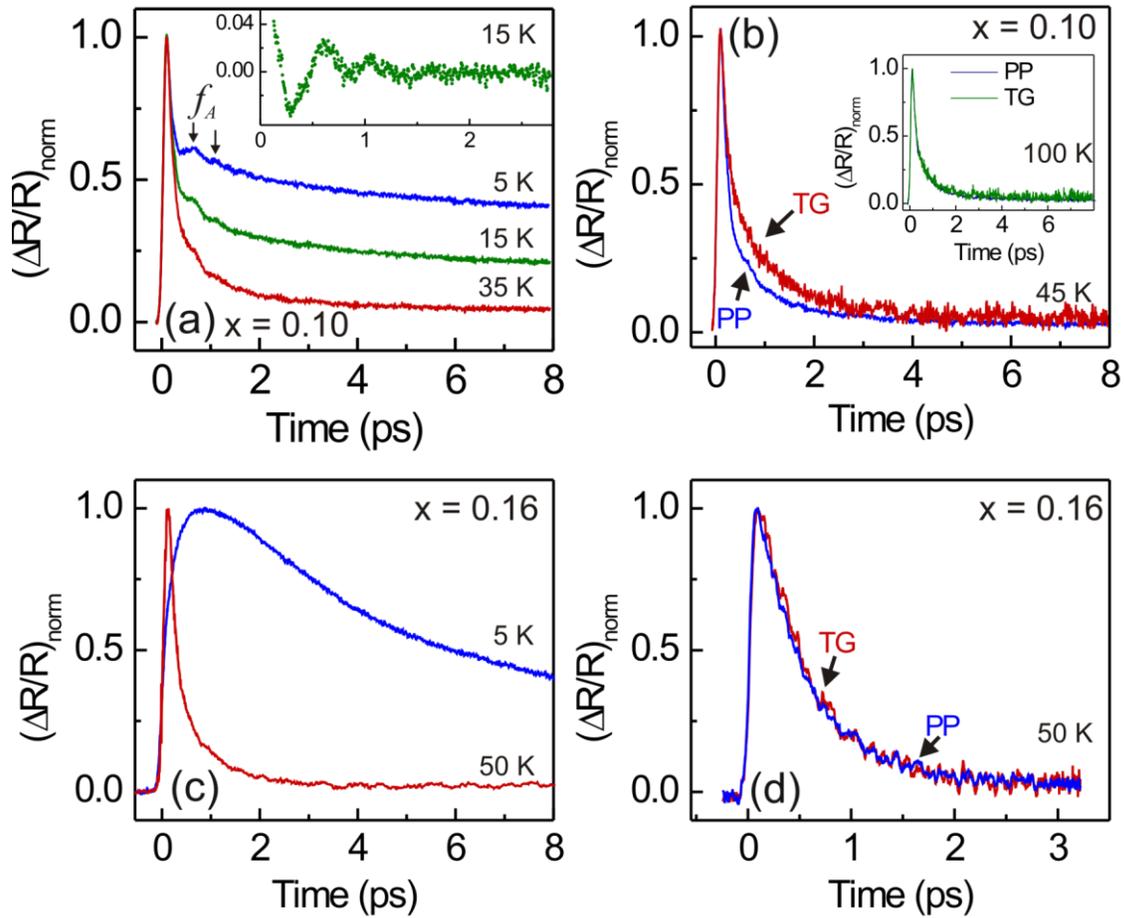

**Figure 2.** Raw data traces indicate the presence of coherent modes of the CDW in La$_{2-x}$Sr$_x$CuO$_4$ for $x = 0.10$ but not for $x = 0.16$. In **(a)**, the amplitudon response is evident via an oscillation of frequency $f_A$ in the reflectivity transients. This response is observed to persist above the superconducting transition temperature $T_c = 26$ K. Inset: the amplitudon response is isolated by subtracting the fitted background electronic response (see text). **(b)** Above $T_c$, there is a discrepancy between the PP and TG responses which we attribute to the phason, as described in the text. This difference disappears at ~ 100 K (inset). **(c)** Data from the $x = 0.16$ sample ($T_c = 38.5$ K) show only the quasiparticle recombination dynamics associated with non-equilibrium excitation of the superconducting state below $T_c$ and a fast electronic transient above $T_c$. There is no evidence of either the amplitudon or phason above or below the superconducting transition temperature. Similar behavior was observed in the $x = 0.36$ sample (not shown). **(d)** TG and PP transients at 50 K in the $x = 0.16$ sample. There is no discernible difference between the two traces.

**Figure 3**

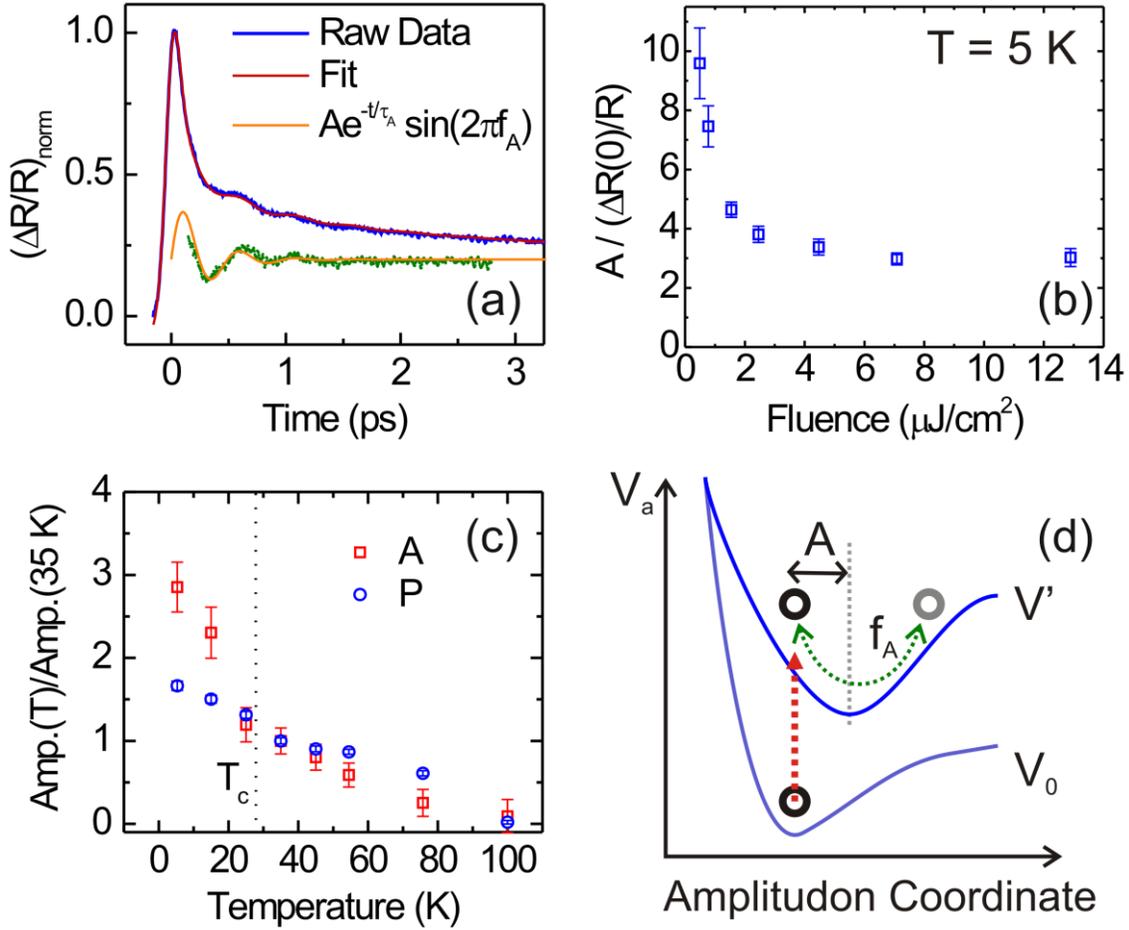

**Figure 3.** Analysis of the amplitudon response as a function of temperature. **(a)** Representative data at $T = 15$ K are shown along with the fit to the model described in the text and the residual due to the amplitudon response. We found excellent agreement between the model function and the data at all temperatures. **(b)** Ratio of the amplitude ($A$) of the amplitudon to the total amount of signal ($\Delta R(0)/R$) as a function of fluence. The saturation of the amplitude with fluence is consistent with other measurements of the amplitudon in conventional CDW systems. **(c)** The magnitude of the amplitudon and phason as a function of temperature, both normalized to the value at 35 K for comparison. Their common disappearance indicates that $T_{CDW} \sim 100$ K. **(d)** Representation of the displacive excitation mechanism (DECP) which drives the amplitudon. The system is represented by an open circle. The abscissa $x$ represents the degree of charge modulation while $V_o$ and $V'$ represent the potential energy surfaces before and after the absorption of laser light, respectively.

**Figure 4**

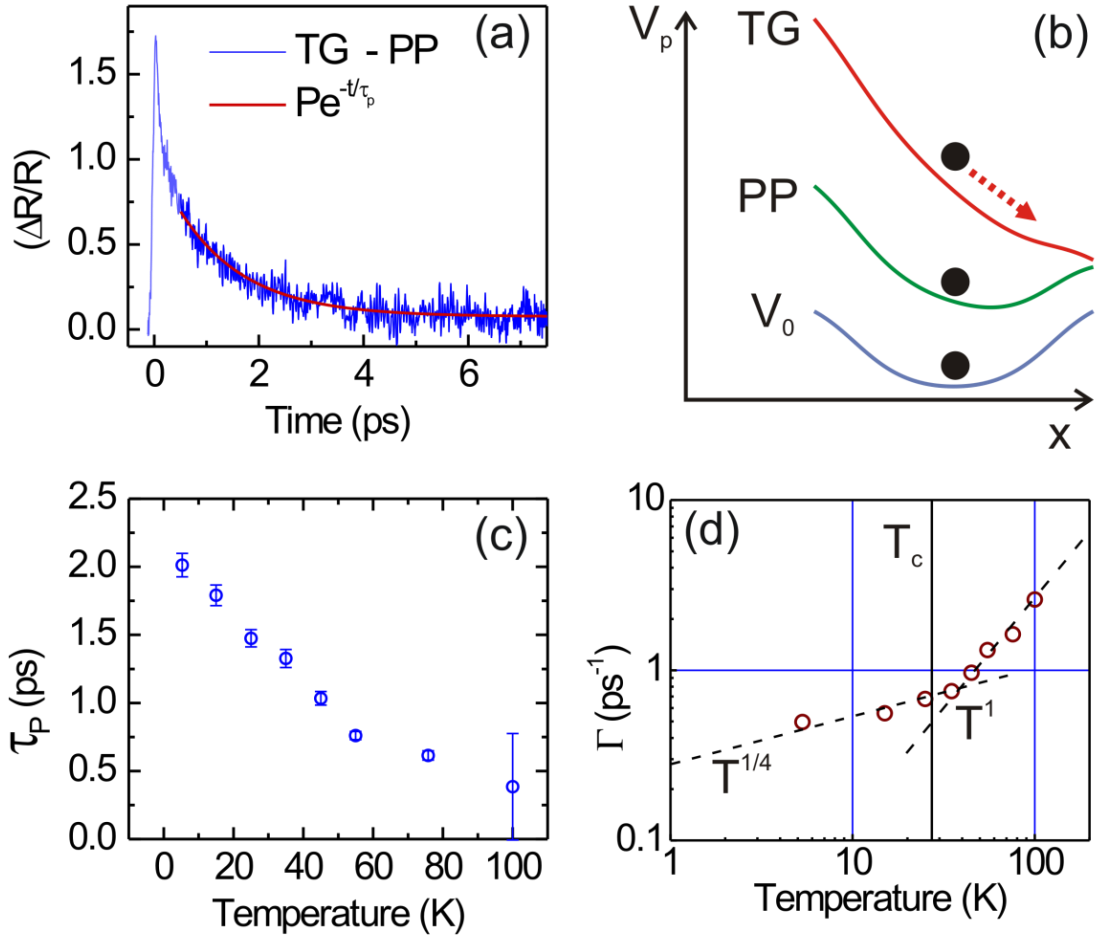

**Figure 4.** Analysis of the phason response as a function of temperature. **(a)** The difference between the TG and PP components yields the pure phason response, which is fit to a decaying exponential yielding the phason magnitude $P$ and lifetime $\tau_P$. **(b)** Illustration of the depinning mechanism for phason generation. The gradient of the TG geometry is ~ 100 times larger than that of the PP geometry, allowing the CDW to slide. **(c)** The CDW fluctuation lifetime decreases with temperature until it becomes immeasurably short at $T_{CDW}$. **(d)** A log-log plot of the phason decay rate as a function of temperature does not show the $\Gamma_P \propto T^5$ behavior expected from intrinsic damping of the phason only.